\documentclass[11.5pt]{scrartcl}

\usepackage{xy-format}
\usepackage{xy-theorem}
\usepackage{xy-math}
\usepackage{fancyhdr}
\usepackage{longtable}
\usepackage{tikz}
\usetikzlibrary{arrows.meta,positioning}
\usepackage{placeins}

\definecolor{sqblue}{HTML}{286284}
\definecolor{sqamber}{HTML}{AC7426}
\tikzset{sqdiagram/.style={font=\sffamily\small,
box/.style={draw=sqblue,fill=sqblue!6,rounded corners=3pt,line width=.7pt,align=center,text width=4cm,minimum height=1.1cm,inner sep=7pt},
wide/.style={box,text width=10.5cm},
private/.style={box,draw=sqamber,fill=sqamber!10},
external/.style={box,draw=gray,dashed,fill=gray!5},
arr/.style={-{Stealth[length=2.2mm]},draw=sqblue!85!black,line width=.8pt}}}

\usepackage{array,listings,xurl,needspace}
\AddToHook{cmd/section/before}{\Needspace{6\baselineskip}}
\hypersetup{pdftitle={StringQFT: A Multiuse Textbook Resource for AI-Assisted Theoretical Physics},pdfauthor={Xingyang Yu}}
\title{\resizebox{\textwidth}{!}{StringQFT: A Multiuse Textbook Resource for AI-Assisted Theoretical Physics}}
\author[1,2]{Xingyang Yu}
\affil[1]{FirstPrinciples, Toronto, Ontario, Canada}
\affil[2]{Department of Physics, Virginia Tech, Blacksburg, Virginia 24061, USA}
\date{}
\begin{document}
\maketitle
\begin{abstract}
We present StringQFT as a textbook collection for string constructions of quantum field theory and as infrastructure for AI-assisted theoretical physics. The collection comprises 18 volumes, 190 chapters and 3,795 pages, covering brane probes, brane webs, geometric engineering, F-theory constructions, wrapped M5-branes and exact worldsheet constructions. Its open-book benchmark inventory contains 1,073 items, including 499 with automatic finite-answer checking. For retrieval-augmented generation, it provides 1,371 section records linked to textbook sources and PDF locations, together with a local lexical retriever. Training-data preparation is illustrated by 21 private, source-linked examples. For agents, it exposes 19 calculation tools with specified inputs and finite domains, connecting tool use to textbook derivations. We illustrate these uses through specific questions, retrieved passages and finite calculations. The textbook collection and accompanying resources are available at \url{https://github.com/xingyang-yu/StringQFT}.
\end{abstract}
\newpage
\begingroup
\small
\setlength{\parskip}{0pt}
\tableofcontents
\endgroup
\newpage
\section{Introduction and Summary}

Over the past several decades, string theory has proved a powerful framework for charting quantum field theories and developing calculational approaches to their dynamics. This perspective draws on many ingredients of string theory, and advances in these ingredients have continually uncovered new structures and results in quantum field theory. In particular, the richness of the string-theory toolkit means that a difficult field-theory question can sometimes be translated into calculations in very different string-theoretic settings.

Learning and connecting these constructions is often challenging, even through explicit examples, because they require familiarity with different aspects of string theory. One needs to understand how the degrees of freedom arise, how interactions or observables are calculated within the string-theory framework, and how the results are interpreted in field theory. The same physical question can appear in the language of open strings, geometry or wrapped branes, with different conventions and calculational tools. Detailed reviews develop individual construction routes. Our aim is to bring these routes together with common conventions, explicit derivations and connected reading paths. Providing this integrated resource is one motivation for the textbook introduced in this note.

LLM agents now make it possible for us to undertake a textbook of this scope through human-supervised drafting and development. This also leads to a second motivation, namely to develop the textbook as a resource for AI-assisted work in theoretical physics. As AI becomes part of physicists' everyday learning and reading, textbooks can be organized to support these activities. Consider an agent, chatbot or other AI assistant helping a reader reproduce a calculation from a textbook, lecture notes or review article. It needs to locate the relevant derivation, recover the conventions used, reproduce the calculation and explain the result in its physical setting. Each step draws on a different part of the same exposition. Making passages, exercises and calculations individually accessible allows these steps to be connected, with the textbook providing the explanation that holds them together.

These uses connect to several strands of work on AI-assisted physics. Benchmarks now test theoretical-physics problem solving \cite{TPBench,QFTBench}, tacit reasoning in QFT and string theory \cite{Tacit}, and research tasks spanning several areas of physics \cite{CritPt,PRLBench,PHYRD}. Retrieval-augmented generation connects model responses to retrieved documents \cite{RAG}, while synthetic instruction data \cite{SelfInstruct} and specialized high-energy-theory models such as FeynTune \cite{FeynTune} provide complementary approaches to training. Tool-using agents combine reasoning with external actions \cite{ReAct}; in theoretical physics, TheoPhysEval evaluates autonomous problem solving with literature search and computation \cite{TheoPhysEval}, and DualityCert studies the repair of candidate duality claims using symbolic consistency checks \cite{DualityCert}.\footnote{Related machine-learning applications to supersymmetric gauge theories include tracing Seiberg-duality sequences \cite{HeckmanTrace}, exploring and generating brane configurations through coamoeba \cite{SeongCoamoeba,SeongGenerative}, and diagnosing inconsistent brane tilings \cite{LeeSeongInconsistencies}. These works develop methods for specific physics problems rather than general-purpose language-model benchmarks.}

With these two motivations in mind, we present StringQFT, a textbook collection on string-theoretic constructions of quantum field theory, together with resources for benchmarking, retrieval, training-example preparation and computational tools for agents. The first release contains 18 volumes, 190 chapters and 3,795 PDF pages. It covers several interconnected approaches to constructing quantum field theories, often studied by distinct research communities:
\begin{itemize}
\item \textbf{Branes probing singularities.}\footnote{See \cite{JohnsonPrimer,YamazakiTilings} and references therein.} Field theories arise on branes placed at geometric singularities, with the singularity and brane data determining their matter content and interactions.
\item \textbf{Suspended branes and brane webs.}\footnote{See \cite{GiveonKutasov} and references therein.} Configurations of branes stretched between other branes or joined into webs encode gauge sectors, matter and parameters through their arrangement and charges.
\item \textbf{Geometric engineering.}\footnote{See \cite{KlemmGeometry} and references therein.} Local geometries in type II string theory and M-theory realize field theories whose light states and couplings are described by wrapped branes, cycles and geometric moduli.
\item \textbf{F-theory constructions.}\footnote{See \cite{Weigand,HeckmanRudelius} and references therein.} Elliptically fibered geometries encode seven-brane configurations and provide constructions of four- and six-dimensional theories through their singularities and additional background data.
\item \textbf{Wrapped M5-branes.}\footnote{See \cite{TachikawaPedestrians,TachikawaTN} and references therein.} Compactifying M5-brane systems on internal manifolds, with suitable twists and defect data, produces lower-dimensional field theories such as theories of class S.
\item \textbf{Exact worldsheet constructions.}\footnote{See \cite{Vonk,MarinoLectures,AdamoTwistor} and references therein.} Worldsheet descriptions, including topological and twistor strings, give direct access to spacetime field theories or selected protected and self-dual sectors.
\end{itemize}

Across these approaches, worked examples explain how the field-theory data follow from the string construction, while exercises ask readers to apply the methods and interpret the results. Readers can study the volumes sequentially or follow a particular construction or physical question across the collection.

The same derivations and exercises also support the four AI-assisted uses described in this note. The open-book benchmark inventory contains 1,073 items, including 499 with automatic finite-answer checking, 573 requiring manual assessment and one unscored demonstration. For retrieval-augmented generation, 1,371 section records and a local lexical retriever provide access to passages with their textbook locations and citation information. Training-data preparation is illustrated by 21 private examples linking prompts and responses to source material. For agents, 19 calculation tools expose selected computations with specified inputs and finite domains. Each resource connects an answer or calculation to its source derivation, conventions and physical assumptions.

A recurring example from Volume V illustrates how these uses fit together. The Gopakumar--Vafa (GV) multicover relation connects Gromov--Witten (GW) expansion coefficients to protected particle coefficients \cite{GV}. The textbook explains the multicover contributions; benchmark questions ask readers to extract coefficients and interpret them; retrieval supplies the relevant conventions and derivation; prepared training examples pair such tasks with explanations; and calculation tools perform the coefficient transformations. This example follows the same physics from exposition through questions, source retrieval and computation.

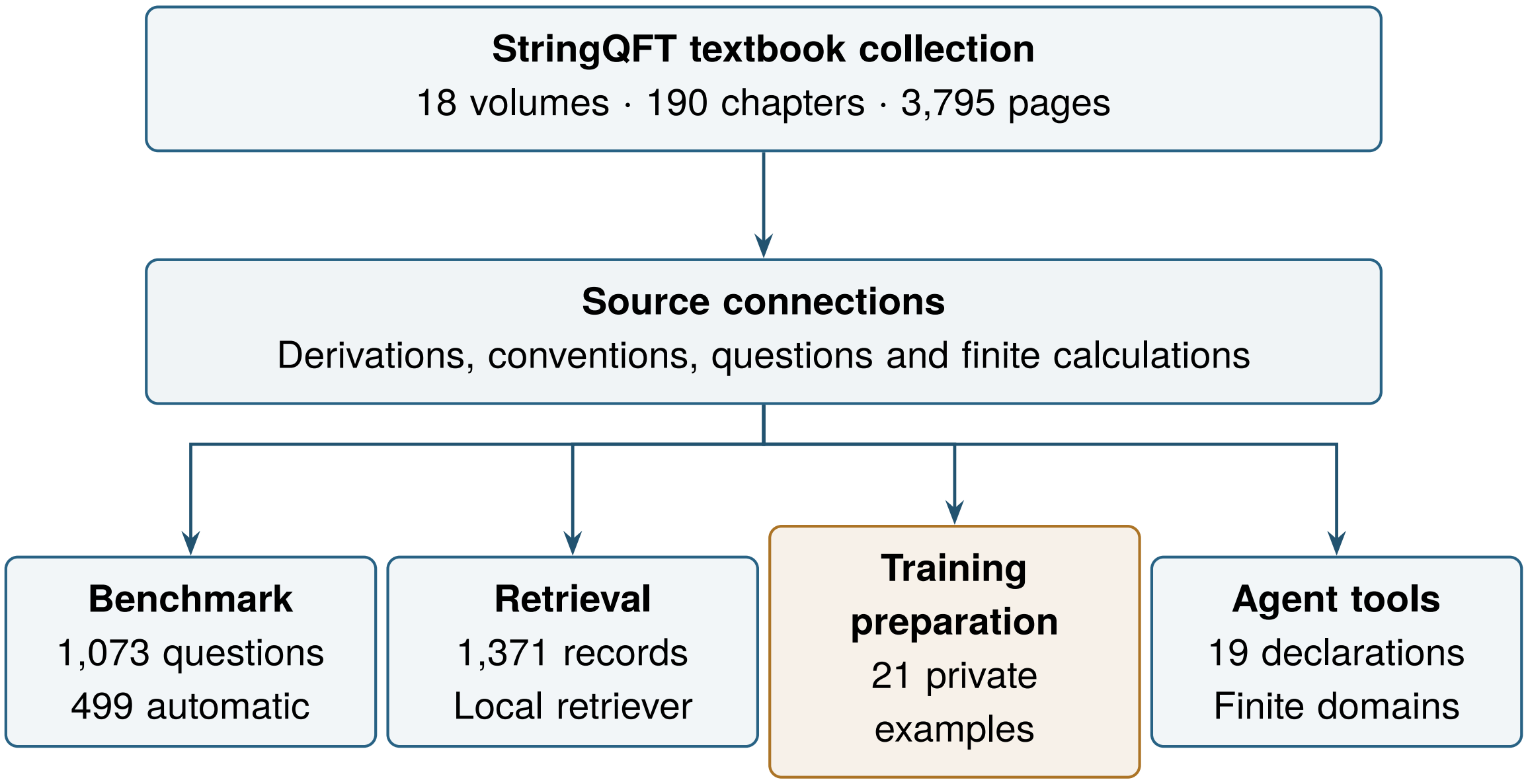
\begin{figure}[!htbp]
\centering
\resizebox{\linewidth}{!}{%
\begin{tikzpicture}[sqdiagram]

\node[wide] (book) at (0,0) {\textbf{StringQFT textbook collection}\\18 volumes $\cdot$ 190 chapters $\cdot$ 3,795 pages};
\node[wide,fill=sqblue!7] (link) at (0,-2.25) {\textbf{Source connections}\\Derivations, conventions, questions and finite calculations};
\node[box, text width=2.8cm] (b) at (-5.1,-5.1) {\textbf{Benchmark}\\1,073 questions\\499 automatic};
\node[box, text width=2.8cm] (r) at (-1.7,-5.1) {\textbf{Retrieval}\\1,371 records\\Local retriever};
\node[private,text width=2.8cm] (t) at (1.7,-5.1) {\textbf{Training preparation}\\21 private examples};
\node[box,text width=2.8cm] (a) at (5.1,-5.1) {\textbf{Agent tools}\\19 declarations\\Finite domains};
\draw[arr] (book)--(link);
\foreach \n in {b,r,t,a} {\draw[arr] (link.south) -- ++(0,-.35) -| (\n.north);}
\end{tikzpicture}}
\caption{The same textbook material supports four uses. Source identifiers and conventions connect questions, retrieved passages, prepared training examples and calculation tools. The amber training block denotes private prepared material; the diagram describes resources, not measured model performance.}
\label{fig:overview}
\end{figure}

Figure~\ref{fig:overview} summarizes these connections.

Readers can browse and download the individual textbook PDFs at \url{https://xingyangyu.com/textbook/}. The sources, code and accompanying resources are available at \url{https://github.com/xingyang-yu/StringQFT} under the release tag \nolinkurl{v1.0}. The public package includes the textbook volumes and sources, retrieval records, calculation interfaces, questions and selected worked examples. Complete solutions, private evaluator inputs and prepared training examples are not included. Licensing scopes and reuse terms are documented in the repository.

The remainder of this note follows the four uses above after introducing the textbook itself. Section 2 describes the collection and its reading routes. Section 3 presents the open-book benchmark and its evaluation procedures. Section 4 describes the retrieval corpus and passage citations. Section 5 discusses the preparation of training examples. Section 6 presents the calculation tools for agents and demonstrates their use with a retrieved passage.
\section{As a textbook}

\subsection{Learning across construction languages}

String constructions of quantum field theory are often learned through a
particular approach, such as brane probes, intersecting branes, geometric
engineering, F-theory or worldsheet models. A reader familiar with one
approach may find it difficult to recognize the physical content of another.
We use the textbook to teach the dictionaries that connect these
descriptions. The aim is to let readers identify the relevant degrees of
freedom, specify the field-theory limit and follow physical quantities
between descriptions with their regimes and conventions stated.

The distinction between open- and closed-string sectors helps explain these
different starting points. In many perturbative D-brane constructions, light
open strings on a stack supply gauge fields, while strings joining different
stacks supply charged matter. Geometric constructions organize the problem
through the background geometry and its deformations, including moduli
described by closed-string fields. The geometry also contains cycles on
which branes can wrap to produce particles or strings. These states can be
essential to the field theory near a singular limit. A reader moving from a
brane description to a geometric one must therefore track both the
background data and the states that become light.

Worldsheet methods help explain both viewpoints. Boundary conditions and
projections determine the open-string modes used in a brane construction.
Topologically twisted worldsheet theories calculate selected protected
quantities that admit a spacetime interpretation. F-theory adds an elliptic
description of the IIB coupling and seven-brane charge transport; its
perturbative D-brane interpretation requires an appropriate weak-coupling
limit. Learning these relations involves identifying what each description
calculates and which physical assumptions connect it to the others.

The pedagogical goal is a coherent understanding of the constructions and
their connections. Readers learn to specify the target theory or sector,
identify a regime in which the description applies and use the relevant
dictionary to compare physical quantities. The scope of a comparison follows
the physical justification given for that dictionary.

\subsection{Common physical questions}

A useful way to approach an unfamiliar construction is to ask which objects
remain light and why they interact as a field theory. A stretched string
may become light when branes approach one another. A wrapped-brane state
may become light when the volume of its supporting cycle decreases. The
next question is which scales can be separated so that the desired sector
survives. For a QFT limit, one asks how massive string modes, unwanted
compactification modes and gravity decouple. The little string limit
instead retains a finite string scale while decoupling gravity. Distinguishing these limits is part of learning what the
construction produces. Within the chosen limit, the dictionary specifies
which brane positions or geometric moduli represent masses and couplings.

We carry these questions across the construction chapters. They provide a
common way to read descriptions whose natural variables differ.

\par\medskip\noindent\begin{minipage}{\linewidth}\small\setlength{\tabcolsep}{5pt}\renewcommand{\arraystretch}{1.2}
\begin{tabular}{@{}>{\raggedright\arraybackslash}p{0.32\linewidth}>{\raggedright\arraybackslash}p{0.62\linewidth}@{}}
\toprule
\textbf{Physical question} & \textbf{Information to identify in the construction} \\
\midrule
Which degrees of freedom survive? & Light modes and wrapped states, with their charges and localization properties \\
Which theory or sector is obtained? & Spacetime dimension, preserved supersymmetry and the field-theory target \\
Which limit defines it? & The scales held fixed, scales removed and assumptions needed for decoupling \\
What are its parameters? & The map from brane or geometric data to masses, couplings and vacuum expectation values \\
What can be compared? & A specified spectrum, moduli space, anomaly or protected coupling in matching conventions \\
\bottomrule
\end{tabular}
\end{minipage}\par\medskip

Conventions enter these questions physically. The orientation of an
intersection affects a signed chiral index. A change of charge basis changes
the components used to describe a state. A rescaling of the formal coupling
changes expansion coefficients. The reader needs the corresponding
translation before comparing answers. Where a result is transferred through
a duality or a protected-sector identification, its physical premises remain
part of that translation.

\subsection{Subject progression and reading routes}

Public volume numbers I--XVIII identify the 18 volumes in this edition. The
following seven blocks describe its subject progression. The intended reader
has a graduate background in QFT; the collection supplies the relevant
supersymmetric toolkit and string dictionaries.

\subsubsection{Foundations}

Volumes I--V establish the common language used by the construction chapters.
Volume I introduces supersymmetric field theories across dimensions.
Volumes II and III develop physical strings and branes, including worldsheet
quantization, boundary conditions, effective actions and duality dictionaries.
Volume IV supplies geometric methods, while Volume V introduces topological
strings and their protected quantities. Readers can use these volumes to
learn the foundations in sequence or return to a particular method when a
construction requires it. Their role is to make the assumptions and
conventions behind later calculations accessible from a common starting point.

\subsubsection{Brane probes}

Volumes VI--VIII study field theories supported on branes probing a background.
Volume VI develops D-branes at orbifold singularities, deriving quiver data
from projections and relating vacuum equations to the probed geometry \cite{DouglasMoore}.
Volume VII extends the discussion to toric singularities and brane tilings \cite{FrancoEtAl}.
Volume VIII develops M2-brane probes and three-dimensional SCFTs, including the ABJM construction \cite{ABJM}. This block
teaches how a probe construction translates background data into fields,
interactions and moduli spaces. The M2 volume extends the probe viewpoint
beyond the perturbative open-string description used for D-branes,
introducing the physical input needed in that setting.

\subsubsection{Brane intersections, suspended branes and webs}

Volumes IX--X develop constructions in which the arrangement and interaction
of several branes organize the field theory. Volume IX treats
Hanany--Witten configurations \cite{HananyWitten}, branches of vacua, brane dynamics and the
D4--NS5 M-theory lift. Volume X develops five-brane webs, seven-brane
junctions and magnetic quivers, building on the five-brane web description \cite{Webs}. The local intersecting-brane spectrum is
introduced earlier in Chapter III.6. Together these chapters teach readers
to distinguish local endpoint and intersection data from the additional
charge, boundary and dynamical information required by a suspended-brane
or web construction.

\subsubsection{Geometric engineering}

Volumes XI--XIII develop geometric engineering in type IIA, type IIB and
M-theory, respectively. The geometric-engineering approach \cite{KatzKlemmVafa} organizes field-theory calculations through local string backgrounds. The geometry specifies cycles, deformations and
intersection data that enter the field-theory dictionary. Wrapped-brane
states, periods and scaling limits explain how these data acquire their
physical meaning. The examples include gauge sectors from ALE fibrations,
theories from singular threefolds and five-dimensional prepotentials from
compact surfaces. The teaching objective is to follow the passage from a
geometric calculation to a specified QFT quantity, including the state and
decoupling assumptions required for that passage.

\subsubsection{F-theory}

Volumes XIV--XV introduce F-theory through its elliptic-geometric and
seven-brane dictionaries \cite{VafaF,Weigand}, then develop four- and six-dimensional
constructions. Volume XIV treats gauge and matter data, fluxes, global
structure and effective-action questions, including the weak-coupling
comparison. Volume XV develops tensor branches, anomalies, SCFT limits
and little string theories. Readers learn how the elliptic description
encodes physical data and which additional background and limiting choices
are needed to specify a theory. This also makes the relation to perturbative
IIB constructions explicit in the regimes where that comparison applies.

\subsubsection{Wrapped M5-branes}

Volumes XVI--XVII approach lower-dimensional theories through compactification
of M5-brane systems. Volume XVI develops the six-dimensional parent and
class S \cite{Gaiotto}, including twists on curves, punctures, Hitchin systems,
Seiberg--Witten geometry and global forms. Volume XVII follows protected
structures and further compactifications, including spectral networks,
indices, wrapped curves with four supercharges, three-manifolds and
divisors. This block teaches how compactification, twisting and boundary
or puncture data determine the lower-dimensional construction. It connects
brane origins with geometric and protected descriptions while stating the
physical input used by each route.

\subsubsection{Worldsheet constructions}

Volume XVIII asks how a worldsheet theory produces a spacetime field theory
or computes a specified sector of it. It begins with topological open-string
reduction and Chern--Simons theory \cite{WittenCS}, then develops open and closed B-models,
matrix-model relations and further worldsheet constructions. The examples
include self-dual theories, twistor strings and two-dimensional Yang--Mills.
Readers follow the relevant worldsheet fields, observables and reductions
into their spacetime interpretation. The physical content and scope of each
example are identified separately, including when a construction describes
a protected or self-dual sector.

\begin{figure}[!htbp]
\centering
\resizebox{\linewidth}{!}{%
\begin{tikzpicture}[sqdiagram,
family/.style={box,text width=3.5cm},
bridge/.style={draw=sqblue,line width=1pt},
shared/.style={draw=sqblue!30,line width=.7pt},
tag/.style={font=\scriptsize,fill=white,inner sep=3pt,align=center}]
\node[family] (probe) at (0,4.8) {\textbf{Brane probes, VI--VIII}\\Orbifolds, tilings\\and M2-branes};
\node[family] (web) at (5.7,2.4) {\textbf{Brane systems, IX--X}\\Suspended branes\\and webs};
\node[family] (m5) at (5.7,-2.4) {\textbf{M5-branes, XVI--XVII}\\Wrapped branes\\and class S};
\node[family] (ft) at (0,-4.8) {\textbf{F-theory, XIV--XV}\\Elliptic geometry\\and seven-branes};
\node[family] (geo) at (-5.7,-2.4) {\textbf{Engineering, XI--XIII}\\IIA, IIB and M-theory\\on local geometries};
\node[family] (ws) at (-5.7,2.4) {\textbf{Worldsheets, XVIII}\\Spacetime theories\\and protected sectors};
\node[box,text width=3.7cm,fill=sqblue!12] (common) at (0,0) {\textbf{Shared physics}\\QFTs, moduli spaces\\and protected observables};
\foreach \n in {probe,web,m5,ft,geo,ws} {\draw[shared] (\n)--(common);}
\draw[bridge] (probe.east) to[bend left=18] node[tag,above,sloped]{T-duality: selected\\probe / brane systems} (web.north);
\draw[bridge] (web.south)--node[tag]{IIA/M-theory lift\\D4--NS5 to M5}(m5.north);
\draw[bridge] (ft.west) to[bend left=18] node[tag,below,sloped]{M/F-theory relation\\circle compactification}(geo.south);
\draw[bridge] (geo.north)--node[tag]{Topological strings\\protected amplitudes}(ws.south);
\node[wide] at (0,-7.4) {\textbf{Common foundations, I--V}\\Supersymmetric QFT, strings, branes, geometry and topological strings};
\end{tikzpicture}}
\caption{Relations among string constructions. Dark links select concrete connections: T-dual probe and suspended-brane descriptions, the D4--NS5 lift to M5-branes, the relation between circle-compactified F-theory and M-theory on an elliptic geometry, and topological-string computations of protected quantities in geometric engineering. These relations require the appropriate backgrounds and limits; they do not identify entire construction families. Pale spokes indicate shared types of physical objects, not a claim that all six descriptions realize the same theory. The foundations supply the common language.}
\label{fig:book_routes}
\end{figure}
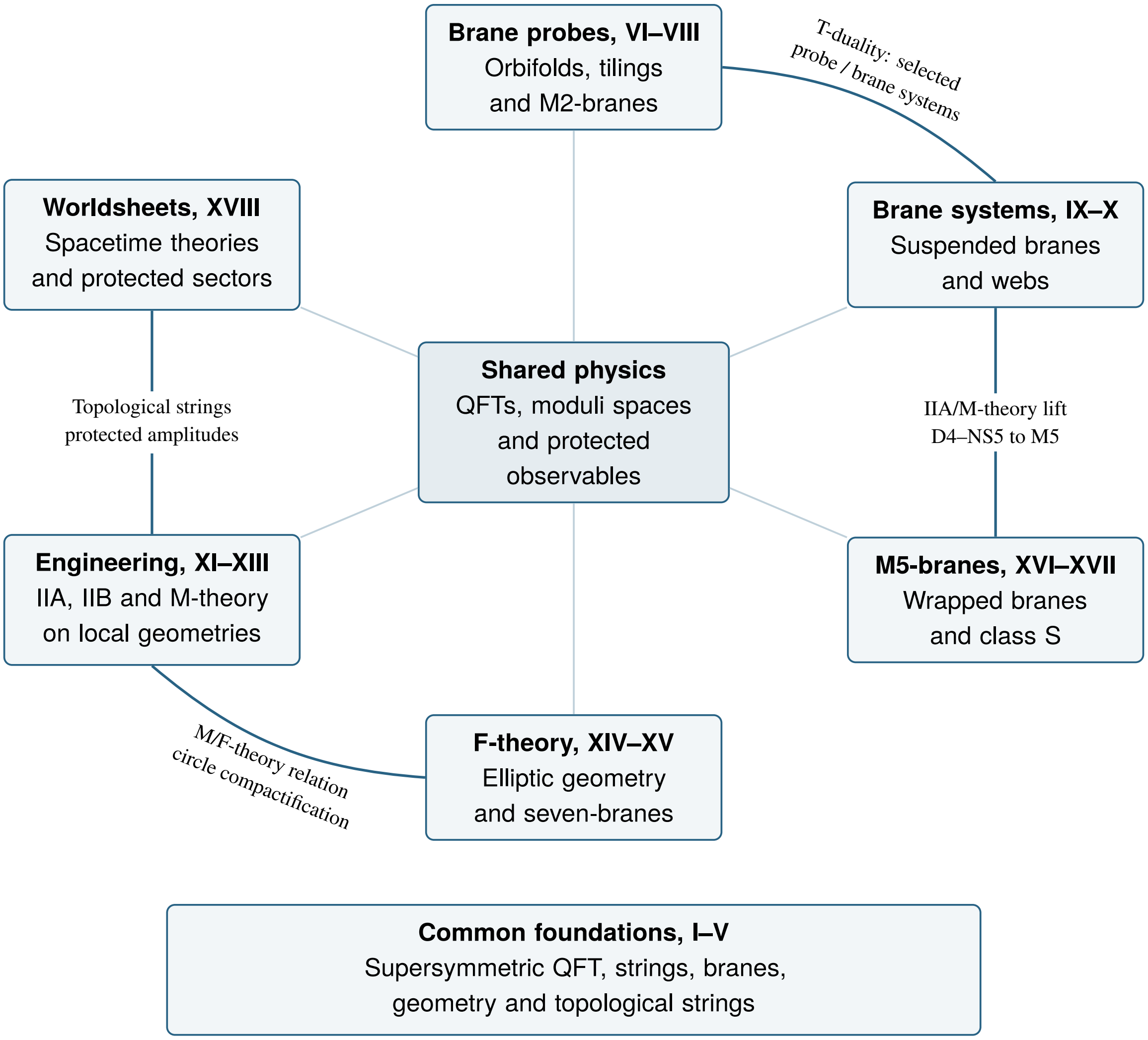

Figure~\ref{fig:book_routes} shows how the construction families connect through selected dualities, limits and descriptions of protected quantities.

\subsubsection*{Reading between the blocks}

Readers can enter through a familiar construction and follow a particular
physical question. For a brane-to-geometry route, Chapters VI.3--VI.4 begin
with an orbifold quiver and recover the geometry of the probe's Higgs branch
from its vacuum equations. The simplest example gives the singular surface
\(uv=w^2\) and, at a nonzero real Fayet--Iliopoulos level with the complex
level zero, its resolution
\(T^*\mathbb P^1\). Solving the field equations and taking the quotient
make the relation between quiver fields and geometric coordinates explicit.
The mathematical identification of the ALE metric is supplied separately
with its hypotheses.

Intersecting branes provide another route into worldsheet reasoning.
Chapter III.6 starts with an open string whose endpoints lie on branes at
different angles. The boundary conditions shift oscillator frequencies;
the stated RNS sector and projection then determine the physical states.
The chapter follows the T-dual description in terms of magnetized branes,
tracking how wrapping numbers become bundle rank and first Chern number.
This gives a reader a concrete reason to learn the boundary theory and its
orientation conventions.

A reader approaching from spacetime geometry can follow Chapter XI.7,
where worldsheet instanton contributions in a local Calabi--Yau are related
to gauge instanton terms in an engineered theory. The calculation holds
the base degree fixed and sums fiber degrees before taking the field-theory
limit. Individual contributions can vanish while the sum remains finite.
The example teaches why the limit and the organization of the expansion
matter when interpreting a worldsheet calculation as a field-theory result.

Wrapped M5-branes connect the suspended-brane and compactification routes.
Chapter IX.5 follows the D4--NS5 configuration into an M5 curve and its
Seiberg--Witten periods under the stated M-theory continuation. Chapters
XVI.2 and XVI.4 approach curve compactifications through the supersymmetry
twist and the Hitchin description. Reading them together helps distinguish
the compactification surface, the spectral curve over a vacuum and the
cycles used to compute periods. Their physical identifications and charge
conventions are specified in the respective chapters.

F-theory has a complementary entry point. Chapter XIV.1 introduces the
elliptic fiber through the varying IIB coupling and the transport of
seven-brane charges. Chapter XIV.9 develops the Sen limit and examines where
the perturbative interpretation applies. For a route from a worldsheet
description to a spacetime action, Chapters XVIII.1--XVIII.2 follow the
topological open-string reduction into Chern--Simons theory, with the quantum
level and framing information introduced separately. These routes let a
reader learn the unfamiliar description through a definite calculation.

These reading routes can also guide a course or independent study.
Instructors can select chapters around a construction or a shared physical
question, combining worked examples with exercises. Readers studying
independently can use the same routes to revisit prerequisites and compare
how different constructions describe related field-theory data.

\section{As a benchmark}

\subsection{Benchmark questions and scoring units}

Scientific benchmarks probe several aspects of model reasoning. GPQA uses expert-written multiple-choice questions in biology, physics and chemistry \cite{GPQA}. Grading the Unspoken examines the reconstruction of tacit reasoning in QFT and string theory \cite{Tacit}. StringQFT provides textbook-linked questions for studying how a model uses supplied derivations, conventions and calculation tools.

The benchmark task is to answer identified questions about string
constructions using the conventions and permitted resources of a specified
textbook edition. The collection is designed for open-book use. Some
derivations and answers appear in the reading text or calculation examples,
so an evaluation must state which books, passages and tools a participant
may consult. The collection supplies question statements, answer formats
and finite checking routes for that task.

The inventory contains 1,073 individually identified questions across all
18 volumes and 190 chapters. Of these, 499 have automatic finite-answer
checking, 573 require manual assessment of written responses and one is an unscored
worked demonstration. A question may request several fields or related
cases; those fields do not become separate scoring units. Chapter
batches aggregate the same individual questions.

Identifiers such as \nolinkurl{V.8.E2} locate a question within its volume and chapter.
Each record specifies its source, question statement, scoring mode and,
where applicable, a structured answer schema. Some records refer to shared
chapter context or supplementary question text. Those references are part of
the question and must accompany it when selecting material for an evaluation.
Automatic coverage varies by volume. For example, Volume V contains 43
automatic questions, 11 manual questions and the demonstration; several
other volumes are entirely manual. Table~\ref{tab:question-modes} records this distinction.

\begin{table}[htbp]
\centering
\caption{Questions and scoring modes by volume. Chapter batches aggregate these questions and do not add scoring units.}
\label{tab:question-modes}
\begin{minipage}{\linewidth}\small\setlength{\tabcolsep}{5pt}\renewcommand{\arraystretch}{1.2}
\begin{tabular}{@{}>{\raggedright\arraybackslash}p{0.16\linewidth}>{\raggedright\arraybackslash}p{0.19\linewidth}>{\raggedright\arraybackslash}p{0.19\linewidth}>{\raggedright\arraybackslash}p{0.19\linewidth}>{\raggedright\arraybackslash}p{0.17\linewidth}@{}}
\toprule
\textbf{Volume} & \textbf{Questions} & \textbf{Automatic} & \textbf{Manual} & \textbf{Demo} \\
\midrule
I & 28 & 0 & 28 & 0 \\
II & 54 & 43 & 11 & 0 \\
III & 54 & 46 & 8 & 0 \\
IV & 62 & 51 & 11 & 0 \\
V & 55 & 43 & 11 & 1 \\
VI & 38 & 0 & 38 & 0 \\
VII & 69 & 0 & 69 & 0 \\
VIII & 56 & 3 & 53 & 0 \\
IX & 35 & 0 & 35 & 0 \\
X & 40 & 32 & 8 & 0 \\
XI & 66 & 0 & 66 & 0 \\
XII & 60 & 4 & 56 & 0 \\
XIII & 81 & 0 & 81 & 0 \\
XIV & 78 & 54 & 24 & 0 \\
XV & 63 & 56 & 7 & 0 \\
XVI & 73 & 73 & 0 & 0 \\
XVII & 78 & 77 & 1 & 0 \\
XVIII & 83 & 17 & 66 & 0 \\
Total & 1073 & 499 & 573 & 1 \\
\bottomrule
\end{tabular}
\end{minipage}
\end{table}

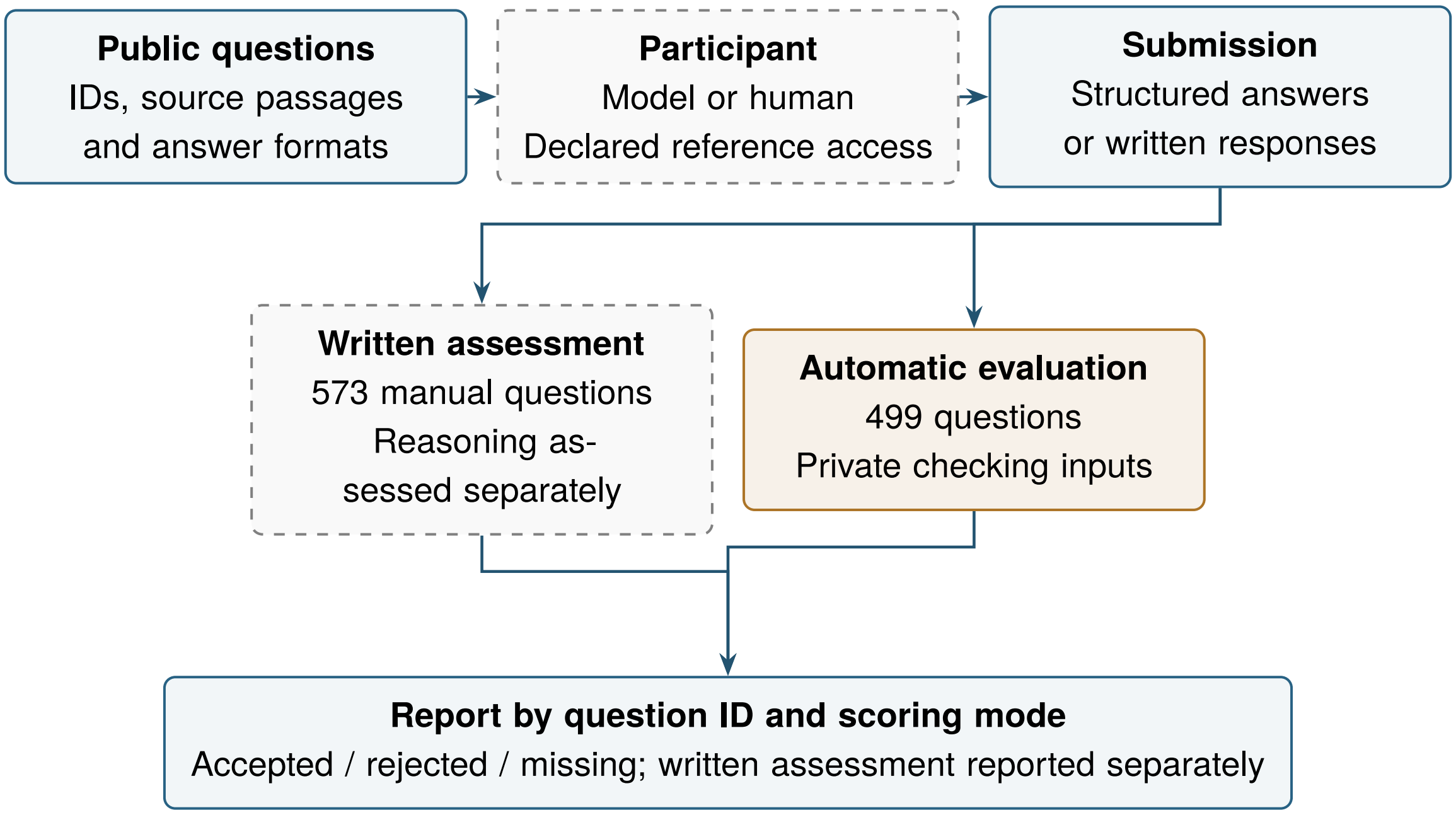
\begin{figure}[!htbp]
\centering
\resizebox{\linewidth}{!}{%
\begin{tikzpicture}[sqdiagram]

\node[box] (q) at (-4.8,0) {\textbf{Public questions}\\IDs, source passages\\and answer formats};
\node[external] (model) at (0,0) {\textbf{Participant}\\Model or human\\Declared reference access};
\node[box] (sub) at (4.8,0) {\textbf{Submission}\\Structured answers\\or written responses};
\node[private] (auto) at (2.4,-3.15) {\textbf{Automatic evaluation}\\499 questions\\Private checking inputs};
\node[external] (manual) at (-2.4,-3.15) {\textbf{Written assessment}\\573 manual questions\\Reasoning assessed separately};
\node[wide] (report) at (0,-6.3) {\textbf{Report by question ID and scoring mode}\\Accepted / rejected / missing; written assessment reported separately};
\draw[arr] (q)--(model);\draw[arr] (model)--(sub);
\draw[arr] (sub.south)-- ++(0,-.35) -| (auto.north);
\draw[arr] (sub.south)-- ++(0,-.35) -| (manual.north);
\draw[arr] (auto.south)-- ++(0,-.35) -| (report.north);
\draw[arr] (manual.south)-- ++(0,-.35) -| (report.north);
\end{tikzpicture}}
\caption{Question selection, submission and assessment. Blue blocks show question, submission and report objects; amber identifies private evaluation material, and dashed blocks denote participants or manual assessment. The one public worked demonstration is excluded from scoring. An automatically accepted answer does not grade its explanation.}
\label{fig:benchmark}
\end{figure}

Figure~\ref{fig:benchmark} follows a question through submission and assessment.

\subsection{From a question to a submission}

Question V.8.E2 asks the reader to invert a supplied rational GW table at
unit coupling scale. Its four rows and four columns determine primitive
candidates through the stated genus and degree. The answer has one required
field, \nolinkurl{primitive}, containing a four-by-four array of rational strings.
The reader must determine the candidates before checking integrality. The
question explicitly treats the input as an algebraic example rather than
asserting that it is a realizable spectrum.

Submissions use a JSON object keyed by question ID. To show a complete
submission without reproducing a private solution, we use the worked
demonstration V.8.E1 supplied in the reader package. It takes the local curve contribution
with one degree-one, genus-zero primitive and writes the formal coupling as
\(\lambda=2u\). The requested table has genus rows zero through three and
degree columns one through four. Its supplied response is

\par\noindent\begin{minipage}{\linewidth}
\begin{lstlisting}
{
  "V.8.E1": {
    "gw": [
      ["1/4", "1/32", "1/108", "1/256"],
      ["1/12", "1/24", "1/36", "1/48"],
      ["1/60", "1/30", "1/20", "1/15"],
      ["1/378", "4/189", "1/14", "32/189"]
    ]
  }
}
\end{lstlisting}
\end{minipage}\par

This example teaches the answer shape and submission envelope. Copying it
earns no point. For an automatically scored question, a participant supplies
the corresponding answer object under that question's ID. An optional
wrapper with \nolinkurl{submission} and \nolinkurl{reasoning} keeps the structured answer
separate from a written explanation. Manual answers are submitted as text.
A chapter batch such as \nolinkurl{V.8.exercises} can contain several individual
question answers; the same question must not also be submitted separately.

\subsection{Scoring and evaluation conditions}

One automatic question contributes one point when all its selected finite
checks pass. Incomplete answer-field groups are rejected; absent questions
remain missing. The evaluator reports \nolinkurl{accepted}, \nolinkurl{rejected} or \nolinkurl{missing}
for automatic questions and aggregates those same units by chapter.
Submitted derivations and reasoning are listed for separate assessment.
Acceptance of a finite answer does not assess the derivation or its physical
interpretation.

The denominator needs particular care for partial submissions. In the
recorded Volume V example, a correct answer to one automatic question gives
one accepted answer out of 43 available automatic questions, with 42
missing. An evaluation should report both the selected question set and
the number actually submitted. That routing example is not a measurement
of a participant's accuracy on the volume.

The public-facing reader package contains the questions, formats and
selected worked examples. Scoring additionally requires the author's
private evaluator, matching native checking programs, control inputs and
dependencies. The public reader alone cannot run those private routes.
The present evaluator is a repository-backed setup; no hosted scoring
service or separately released self-contained evaluator is supplied.

The recorded replay covers all 499 automatic questions with accepted,
incorrect and missing submissions. Representative adverse controls test
selected answer fields, and targeted regressions include a mathematically
legal wrong expression and a modified checker with a deliberately omitted
comparison. The added negative case exposes that modified checker's
incorrect acceptance. These checks
test the stated scoring behavior. They do not exhaust all nested answer
values or independently establish the physical correctness of every
reference answer. A separate fresh-environment replay exercised individual
and chapter-batch routing, a reasoning wrapper and exclusion of the public
demonstration using separately supplied private inputs.

For an evaluation, record the resource version and PDF hashes, selected
question IDs, scoring modes, permitted references and tool access.
Distinguish finite-answer results from human assessment of explanations.
Because textbook passages and examples overlap exercise answers, results
obtained with those resources measure the declared open-book task.
Training on overlapping material requires a separately justified evaluation
selection. The collection currently provides neither a model leaderboard
nor a measured comparison of model capabilities.
\section{As a retrieval resource}

Retrieval-augmented generation supplies retrieved documents as context for a language model \cite{RAG}. StringQFT provides a domain-specific corpus and a local lexical retriever for this use, with each returned passage linked to its location in the textbook.

A reader or application often needs a particular convention before using
a formula. A useful retrieval result must therefore return enough context
to inspect the definition and locate its derivation. StringQFT supplies
1,371 numbered section records containing textbook source text and citation
information. The supplied local retriever ranks these records and returns
the selected passages with their metadata.

A record identifies its volume and section, title and teaching-text edition.
It includes the section's TeX, equation and figure labels, a source path and
source hash, together with the matching PDF destination and page. These
fields connect the returned text to a particular edition of the reading
material. They also allow an application to keep equation labels and
convention statements attached to its source citation.

Each record retains a section's explanation around its formulas, including
such information as expansion parameters, summation ranges and index
conventions. Definitions introduced elsewhere may still require consulting
the surrounding chapter through the source citation.

The Volume V retriever uses local lexical ranking. It tokenizes the query
and source text, weights query-term matches by their frequency across the
section collection and gives additional weight to title matches. It requires
no model service or embedding index. A query with no matching terms can
return no passage; a matching result still needs to be read in context.
The supplied method is a concrete way to access the records, and other
retrievers can use the same section identifiers and citations.

Different names for the same physical object need not share query terms,
and matching terminology does not ensure that the physical assumptions
agree. An application should inspect the returned passage and retain its
source information when using it to construct an answer.

For example, record \nolinkurl{V.8.3} is titled \emph{The multicover organization}. Its source
is \nolinkurl{latex/ch08_gv_invariants/chapter.tex}, and its labels include the winding
measure and forward coefficient relation used in the agent example. The record's
teaching-text edition is \nolinkurl{stringqft-1.0/V}. Source and PDF hashes identify the matching files.

From the Volume V directory, the example query is

\par\noindent\begin{minipage}{\linewidth}
\begin{lstlisting}
python rag/retrieve.py "multicover winding sine pole" --limit 1
\end{lstlisting}
\end{minipage}\par

The recorded result is \nolinkurl{V.8.3}. Its PDF citation is
\nolinkurl{StringQFT_Volume_V.pdf#nameddest=section.8.3}, which opens at file page 134
in the selected PDF. The application can present this citation alongside
the returned TeX and let the reader inspect the full derivation. The source
hash identifies the chapter file from which the passage was extracted.

The returned passage can supply context to a downstream answering system.
For the multicover example, that context contains the positive-degree
restriction, the coupling frame and the winding calculation. An application
can then explain a coefficient or propose a tool call while keeping the
textbook reference visible. 

The retriever itself returns passages rather than generated answers. The
recorded example establishes that the documented query retrieves the named
passage and that its citation resolves in the matching PDF. Identifier and
anchor checks support source navigation. These checks do not measure
retrieval quality over a representative query set, answer faithfulness or
the performance of a generated-answer system.

\begin{figure}[!htbp]
\centering
\resizebox{\linewidth}{!}{%
\begin{tikzpicture}[sqdiagram]

\node[box] (query) at (-4.8,0) {\textbf{Query}\\\texttt{multicover winding}\\\texttt{sine pole}};
\node[box] (corpus) at (4.8,0) {\textbf{Volume V corpus}\\One of 18 volume corpora};
\node[box] (retrieve) at (0,-2.55) {\textbf{Local lexical retrieval}\\Term weighting\\and title matches};
\node[wide] (result) at (0,-5.1) {\textbf{Returned passage and source destination}\\V.8.3, conventions and derivation; matching PDF page 134};
\node[external,text width=10.5cm] (answer) at (0,-7.8) {\textbf{Downstream answering system}\\Passage supplied as context\\Answer with a source citation};
\draw[arr] (query.south)|-(retrieve.west);\draw[arr] (corpus.south)|-(retrieve.east);
\draw[arr] (retrieve)--(result);\draw[arr,dashed] (result)--(answer);
\end{tikzpicture}}
\caption{Retrieval supplies a passage together with its source location. The demonstrated query searches the Volume V corpus; the 1,371-record total covers all 18 volumes. The dashed answering component is an external application; the supplied retriever itself does not generate an answer.}
\label{fig:retrieval}
\end{figure}
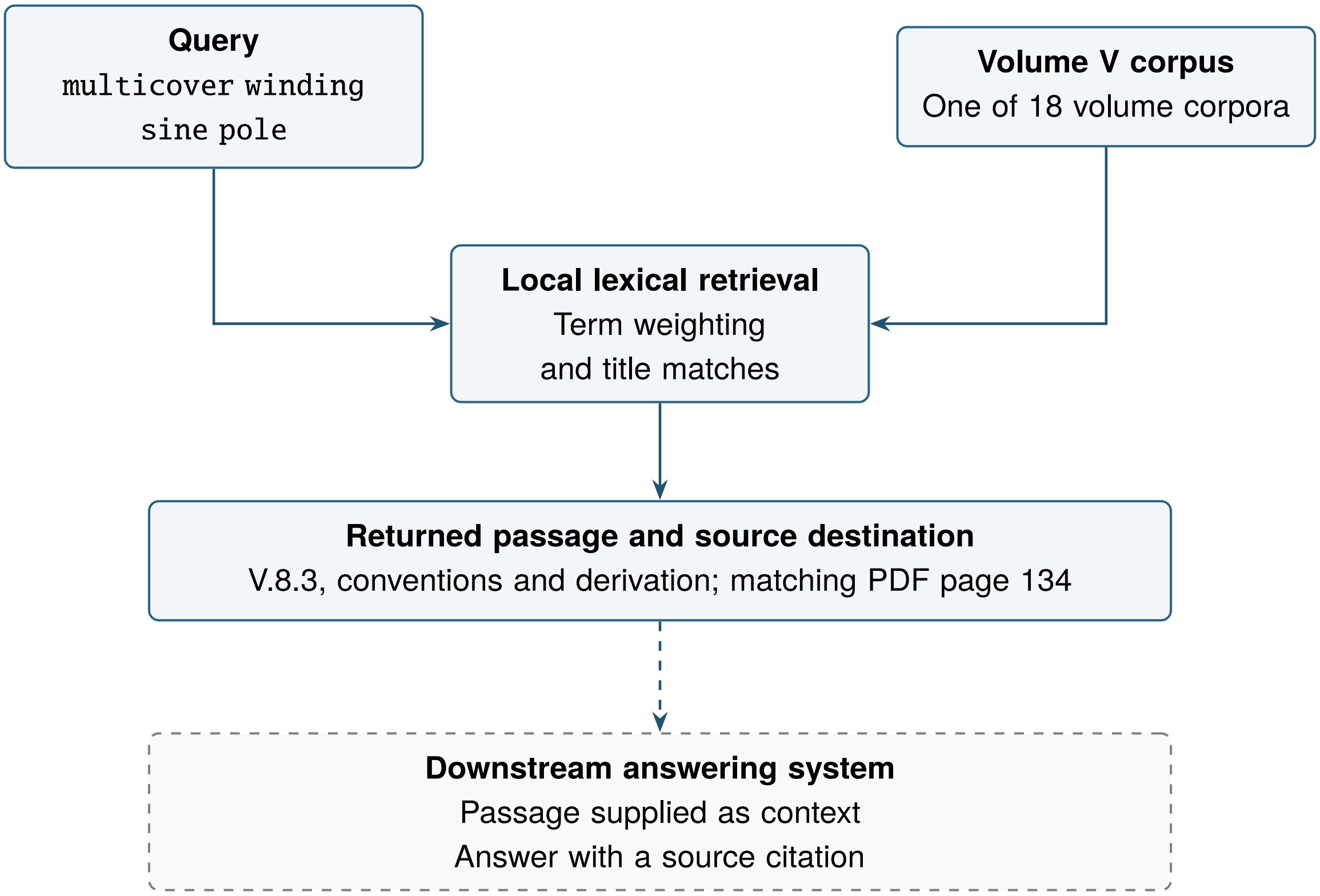

The 18 retrieval corpora are included in the author's selected CC BY 4.0
content scope. That grant covers the author's rights in the selected
expression, with attribution and the applicable third-party exclusions.
Other figures and expression outside the grant retain their separate terms.
An application should keep the matching PDF and versioned source information
with the records. Volume V includes its PDF hash in each passage record;
other volumes bind the PDF through the package and question metadata.
No hosted question-answering service is supplied.
Figure~\ref{fig:retrieval} separates the supplied retrieval operation from a downstream answering application.

\section{As a training resource}

Instruction-tuning work such as Self-Instruct organizes tasks into instructions, inputs and target responses \cite{SelfInstruct}. For textbook-based training, this format can be supplemented with the source derivation and its physical assumptions.

The textbook contains material from which training examples can be prepared,
including explanations of conventions, finite calculations and exercise
responses. The source connections used for reading and retrieval also help
specify what an example teaches. A response can identify the equation it
uses, distinguish a physical premise from its algebraic consequence and
state the domain in which the answer applies.

The current preparation consists of 21 private example records. Four
Volume V examples address the winding measure, a coupling-scale mistake,
triangular inversion and the interpretation of a noninteger result.
Seventeen further records provide a calculation example for each of the
other volumes. These are a small set of prepared examples rather than a
public training collection or evidence of a completed training run.

Each record has an identifier, source information, a prompt, an authored
explanation and an answer, together with its use status. The source fields
vary with the example. The Volume V records identify sections and equation
labels; the other-volume records identify the chapter, calculation-source
hash and PDF citation and hash. The explanation field records an authored
teaching sequence linking the stated premises to the answer.

The numerical Volume V examples were checked against the GV functions, and
the broader staging record reports the prepared calculation examples.
These checks support the finite answers under the stated conventions.
Source references and authored explanations make the premises inspectable;
a numerical check alone cannot certify a physical interpretation or the
quality of a training objective.

Some conceptual examples share subject matter with the open-book questions,
and the textbook contains derivations of exercise results. Training and
evaluation selections must therefore account for this overlap. In particular,
an open-book score on overlapping Volume V material would not establish
transfer to independent questions after training on these examples. No
separate training/evaluation split or contamination-control procedure is
claimed for the current preparation.

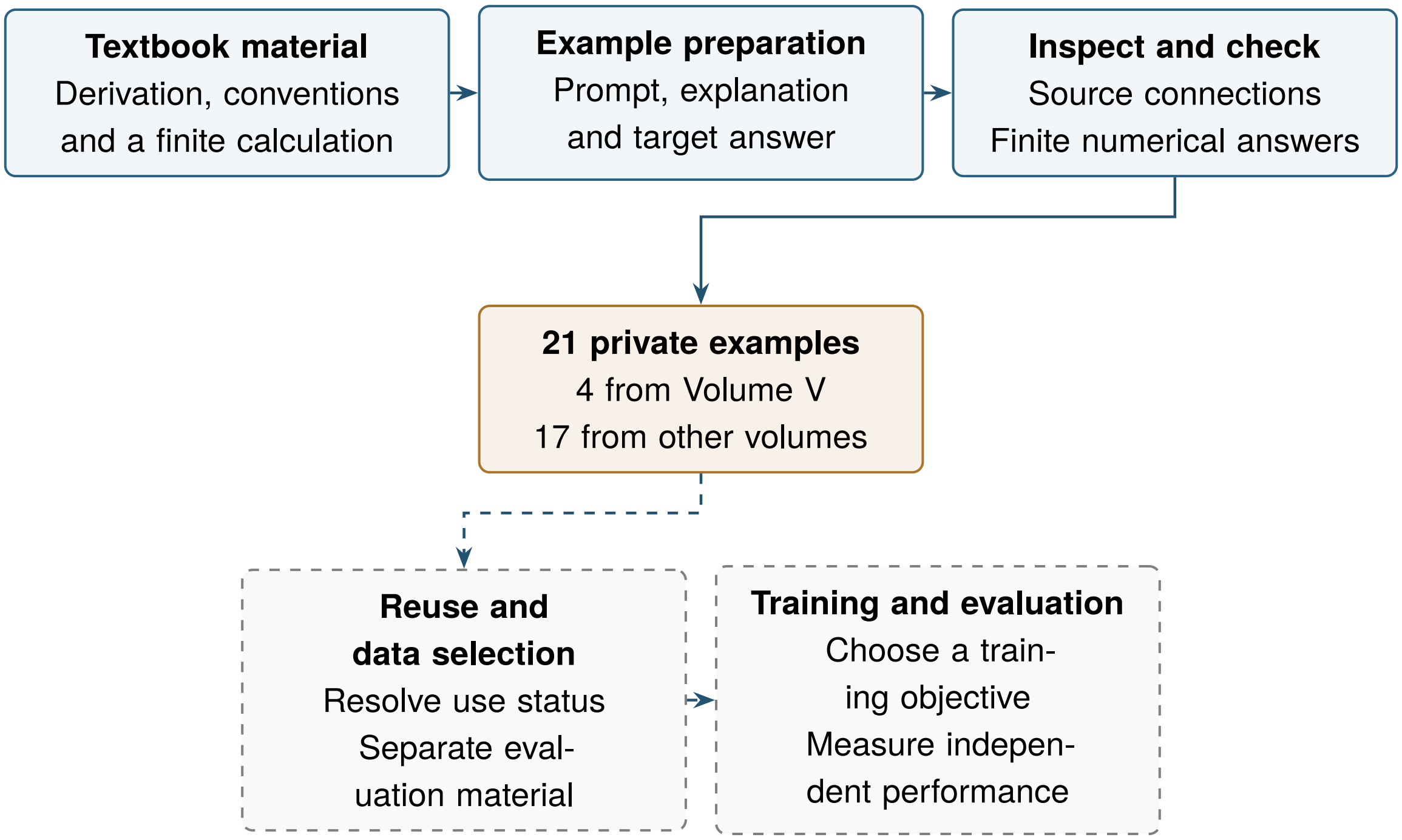
\begin{figure}[!htbp]
\centering
\resizebox{\linewidth}{!}{%
\begin{tikzpicture}[sqdiagram]

\node[box] (source) at (-4.8,0) {\textbf{Textbook material}\\Derivation, conventions\\and a finite calculation};
\node[box] (prepare) at (0,0) {\textbf{Example preparation}\\Prompt, explanation\\and target answer};
\node[box] (check) at (4.8,0) {\textbf{Inspect and check}\\Source connections\\Finite numerical answers};
\node[private] (records) at (0,-3.0) {\textbf{21 private examples}\\4 from Volume V\\17 from other volumes};
\node[external] (selection) at (-2.4,-6.15) {\textbf{Reuse and data selection}\\Resolve use status\\Separate evaluation material};
\node[external] (train) at (2.4,-6.15) {\textbf{Training and evaluation}\\Choose a training objective\\Measure independent performance};
\draw[arr] (source)--(prepare);\draw[arr] (prepare)--(check);
\draw[arr] (check.south)-- ++(0,-.4) -| (records.north);
\draw[arr,dashed] (records.south)-- ++(0,-.4) -| (selection.north);
\draw[arr,dashed] (selection)--(train);
\end{tikzpicture}}
\caption{Training-example preparation and possible subsequent use. The private records pair tasks with authored explanations and source information; numerical examples have finite checks. Dashed blocks identify subsequent reuse, split design, training and evaluation that are not completed experiments in this work.}
\label{fig:training}
\end{figure}

The private examples retain their recorded pending reuse status. The
CC BY 4.0 grant for selected public passages and questions does not extend
automatically to these private records. The current contribution is the
preparation method and its examples; training benefits and generalization
remain unmeasured.
Figure~\ref{fig:training} places the prepared examples within a possible training process.

\section{As a resource for agents}

\subsection{Calculations with declared conventions}

Agent methods such as ReAct interleave reasoning with actions that obtain information from an external environment \cite{ReAct}. Textbook retrieval and callable calculations provide two such actions for a physics assistant.

An agent using a physics textbook needs to know when a calculation is
applicable and how its result should be interpreted. StringQFT exposes
19 declared calculation tools with specified inputs and finite domains.
Their declarations connect callable operations to textbook material. The
agent can retrieve a convention, prepare an input and return both the
calculated result and the passage explaining it.

The declarations and calculation domains vary by volume.
The worked Volume V client is deterministic and can run without a language
model or an API key. It demonstrates a sequence that an external agent can
call, with the physical interpretation supplied by the textbook.

The Volume V example begins by retrieving \nolinkurl{V.8.3} with the query from
Section 4. It then calls \nolinkurl{gv_inverse} on a rectangular coefficient table
in the retrieved convention. The request is

\par\noindent\begin{minipage}{\linewidth}
\begin{lstlisting}
{"scale": "3", "gw": [["5/9", "-11/72"]]}
\end{lstlisting}
\end{minipage}\par

Rows specify genera beginning at zero and columns specify degrees beginning
at one. \nolinkurl{scale} gives the positive factor in the formal coupling
\(\lambda=\mathrm{scale}\,u\). The interface accepts exact integers or
rational strings and rejects floating-point coefficients. In this request
there is one genus row and two degrees. Writing $N_{0,d}$ for the
coefficient of $u^{-2}$ at degree $d$ in this chosen coupling frame, the
input is $N_{0,1}=5/9$, $N_{0,2}=-11/72$. The returned GV coefficients
$n^0_d$ are

\par\noindent\begin{minipage}{\linewidth}
\begin{lstlisting}
[
  {"degree": 1, "genus": 0, "value": "5"},
  {"degree": 2, "genus": 0, "value": "-2"}
]
\end{lstlisting}
\end{minipage}\par

\subsection{Roundtrip verification and use by an agent}

The response reports no noninteger positions and sets \nolinkurl{forward_roundtrip}
to \nolinkurl{true}. The client then sends the primitive entries to \nolinkurl{gv_forward} with
degree two, genus zero and the same scale. The resulting rational table
equals the input. At genus zero the same calculation reads
\[
 n^0_1=3^2\frac{5}{9}=5,\qquad
 n^0_2=3^2\left(-\frac{11}{72}\right)-\frac{5}{8}=-2.
\]
The subtraction removes the double cover of the degree-one contribution.
The negative result is consistent with a signed protected coefficient.

The declaration bounds the total degree by six and the genus by five.
The inverse operation accepts a nonempty rectangular table within those
dimensions. The forward operation specifies its truncation and treats
omitted primitive entries as zero. These bounds define the supplied finite
operation. A successful roundtrip checks consistency of the forward and
inverse transformations; it does not independently validate their common
physical formula or show that arbitrary input coefficients describe a physical BPS spectrum.
That interpretation still needs the protected-state and geometric premises
of the cited chapter.

From the Volume V directory, install the two pinned dependencies and run
the client as follows.

\par\noindent\begin{minipage}{\linewidth}
\begin{lstlisting}
python3 -m venv .venv
. .venv/bin/activate
python -m pip install -r requirements.txt
python agent/example.py
\end{lstlisting}
\end{minipage}\par

The recorded fresh-environment run used macOS with Python 3.9.6, SymPy
1.14.0 and mpmath 1.3.0. It began without SymPy and installed the packages
from PyPI into an isolated environment with no inherited site packages.
Retrieval, the chapter calculation, the inverse/forward client and public
demonstration preparation succeeded. An adverse input using a Boolean
coefficient was rejected. These results establish the documented Volume V
path on that configuration. The other-volume operations retain their own
recorded checks and dependencies; this run did not freshly install them.

\begin{figure}[!htbp]
\centering
\resizebox{\linewidth}{!}{%
\begin{tikzpicture}[sqdiagram]

\node[wide] (retrieve) at (0,0) {\textbf{Retrieve V.8.3 and choose the coupling scale}\\$\lambda=3u$, genus zero, degrees one and two};
\node[box] (input) at (-4.8,-2.85) {\textbf{Exact input}\\$N_{0,1}=5/9$\\$N_{0,2}=-11/72$};
\node[box] (inverse) at (0,-2.85) {\textbf{\texttt{gv\_inverse}}\\$n^0_1=5$\\$n^0_2=-2$};
\node[box] (forward) at (4.8,-2.85) {\textbf{\texttt{gv\_forward}}\\Same scale and truncation\\Reconstruct input};
\node[wide] (compare) at (0,-5.7) {\textbf{Exact roundtrip and source-linked result}\\Compare with the original table; retain the request, convention and citation};
\draw[arr] (retrieve.south)-- ++(0,-.35) -| (input.north);
\draw[arr] (input)--(inverse);\draw[arr] (inverse)--(forward);
\draw[arr] (forward.south)-- ++(0,-.35) -| (compare.north);
\draw[arr,dashed] (input.south)-- (input.south |- compare.north);
\end{tikzpicture}}
\caption{The deterministic Volume V client retrieves a convention, inverts exact coefficients and reconstructs the input. The dashed connection supplies the original table for comparison. This is an available sequence of operations for an external agent, not a measured autonomous reasoning loop or a test of whether the coefficients arise from a physical BPS spectrum.}
\label{fig:agent}
\end{figure}
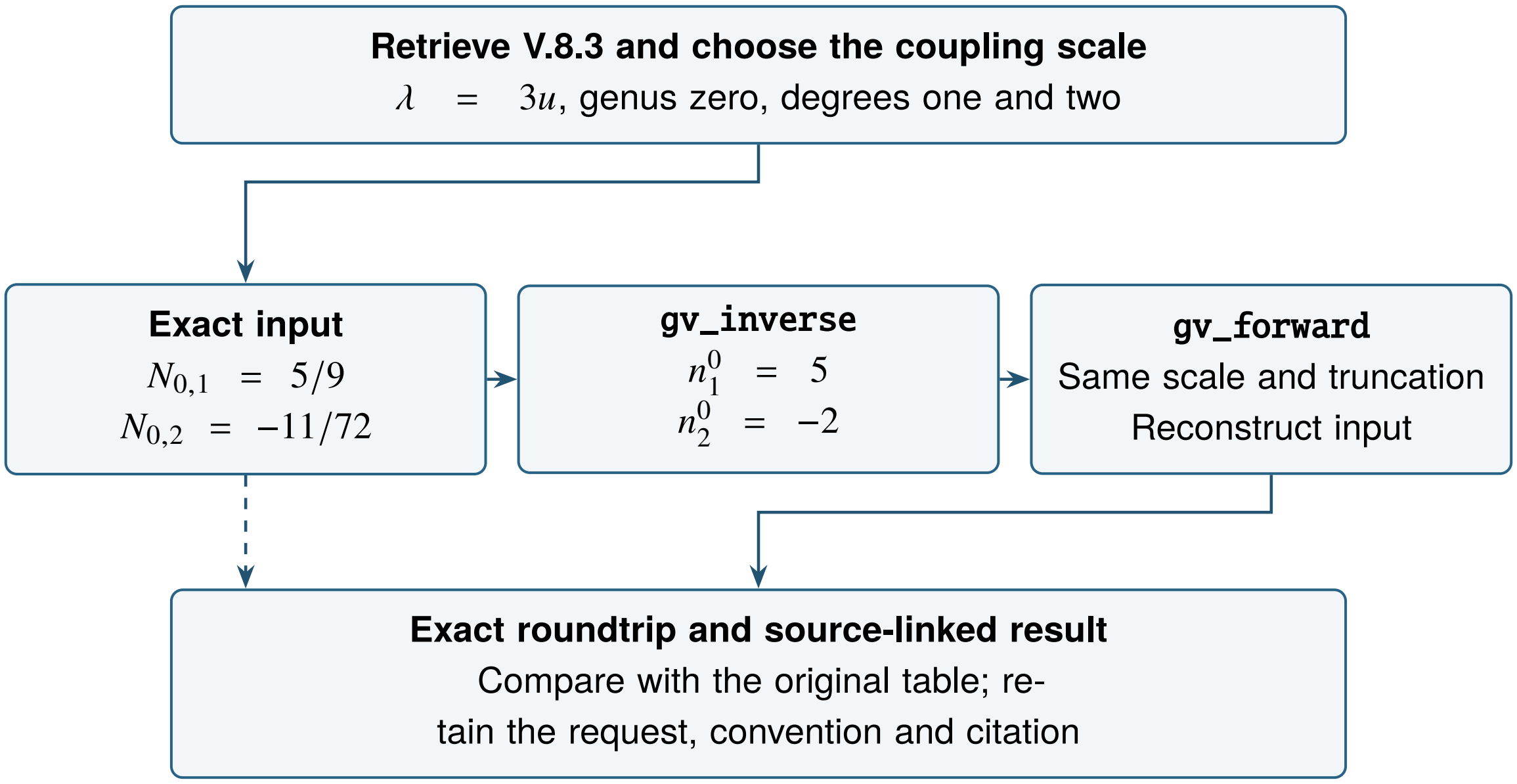

An external agent can use the same sequence while retaining the passage ID,
input convention, exact request and result. If the question lies outside a
tool's declared domain, the interface does not supply an answer for that
broader problem. Selecting a physical model, establishing its assumptions
and judging an interpretation remain separate tasks. The example documents
bounded calculation and source access, with no measured autonomous planning,
learning or research outcome.
Figure~\ref{fig:agent} summarizes the demonstrated retrieval and calculation sequence.

\FloatBarrier
\section*{Acknowledgements}
\addcontentsline{toc}{section}{Acknowledgements}

The textbook, this manuscript and the accompanying code and documentation
were developed with assistance from Anthropic's Claude Opus and OpenAI's
GPT Sol and GPT Astra, under the author's supervision. The author takes responsibility for the
contents and any remaining errors. The author thanks Anthropic for providing a
Claude Max plan during the very early stage of this project.
\par
\phantomsection
\addcontentsline{toc}{section}{References}
\setlength{\bibitemsep}{0.3\baselineskip}
\printbibliography
\end{document}